\documentclass[twocolumn,aps,pra,superscriptaddress]{revtex4-1}
\usepackage{graphicx}
\usepackage{natbib}
\usepackage{url}
\usepackage{textcase}
\usepackage{bm}
\usepackage{color}
\usepackage{times}
\usepackage{amsmath}
\usepackage{physics}
\begin{document}
\title{Fast two-qubit logic with holes in germanium}
\author{N.W. Hendrickx}
\thanks{These two authors contributed equally to this work}
\author{D.P. Franke}
\thanks{These two authors contributed equally to this work}
\author{A. Sammak}
\affiliation{QuTech and Netherlands Organisation for Applied Scientific Research (TNO), Stieltjesweg 1, 2628 CK Delft, The Netherlands}
\author{G. Scappucci}
\author{M. Veldhorst}
\affiliation{QuTech and Kavli Institute of Nanoscience, Delft University of Technology, P.O. Box 5046, 2600 GA Delft, The Netherlands}
\date{\today}

\maketitle

\textbf{The promise of quantum computation with quantum dots has stimulated widespread research. Still, a platform that can combine excellent control with fast and high-fidelity operation is absent. Here, we show single and two-qubit operations based on holes in germanium. A high degree of control over the tunnel coupling and detuning is obtained by exploiting quantum wells with very low disorder and by working in a virtual gate space. Spin-orbit coupling obviates the need for microscopic elements and enables rapid qubit control with Rabi frequencies exceeding 100 MHz and a single-qubit fidelity of 99.3 \%. We demonstrate fast two-qubit CX gates executed within 75 ns and minimize decoherence by operating at the charge symmetry point. Planar germanium thus matured within one year from a material that can host quantum dots to a platform enabling two-qubit logic, positioning itself as a unique material to scale up spin qubits for quantum information.}

Gate-defined quantum dots were recognized early on as a promising platform for quantum information \cite{loss_quantum_1998} and a plethora of materials stacks has been investigated as host material. Initial research mainly focused on the low disorder semiconductor gallium arsenide \cite{koppens_driven_2006, petta_coherent_2005}. Steady progress in the control and understanding of this system culminated in the initial demonstration and optimization of spin qubit operations \cite{bluhm_dephasing_2011,foletti_universal_2009} and the realization of rudimentary analog quantum simulations \cite{hensgens_quantum_2017}.
However, the omnipresent hyperfine interactions in group III-V materials seriously deteriorate the spin coherence, despite attempts to mitigate this by nuclear polarization \cite{bluhm_enhancing_2010}. Drastic improvements to the coherence times could be achieved by switching to the group IV semiconductor silicon, in particular when defining spin qubits in isotopically purified host crystal with vanishing concentrations of nonzero nuclear spin \cite{veldhorst_addressable_2014}. This enabled single qubit rotations with fidelities beyond 99.9\% \cite{yoneda_quantum-dot_2017} and the execution of two-qubit logic gates with fidelities up to 98\% \cite{veldhorst_two-qubit_2015, zajac_resonantly_2018, watson_programmable_2018, huang_fidelity_2018}, underlining the potential of spin qubits for quantum computation. Nevertheless, quantum dots in silicon are often formed at unintended locations and control over the tunnel coupling determining the strength of two-qubit interactions is limited. Moreover, the absence of a sizable spin-orbit coupling for electrons requires the inclusion of microscopic components such as on-chip striplines or nanomagnets close to each qubit, which seriously complicates the design of large and dense 2D-structures. This, combined with the limited control over the location and coupling of the dots, remains an outstanding challenge for the scalability of these systems and a platform that can overcome these limitations would be highly desirable.

Hole states in semiconductors typically exhibit strong spin-orbit coupling, which has enabled the demonstration of fast single qubit rotations \cite{nadj-perge_spinorbit_2010,watzinger_germanium_2018} and additionally, unlike electrons, holes do not suffer from nearby valley states. In silicon, unfavorable band alignment prevents strain engineering of low disorder quantum wells for holes, restricting experiments to metal-oxide-semiconductor (MOS) structures \cite{li_pauli_2015,liles_spin_2018}. Research on germanium mostly focused on self-assembled nanowires \cite{hu_hole_2012,brauns_electric-field_2016,higginbotham_hole_2014} and demonstrated single-shot spin readout \cite{vukusic_single-shot_2018} and coherent spin control \cite{watzinger_germanium_2018}. However, strained germanium quantum wells were recently shown to support the formation of gate-controlled planar hole quantum dots \cite{hendrickx_gate-controlled_2018, hardy_single_2019}. Now, the crucial challenge is the demonstration of coherent control in this platform and the implementation of qubit-qubit gates for quantum information with holes.

\begin{figure*}
	\includegraphics[width=18.5cm]{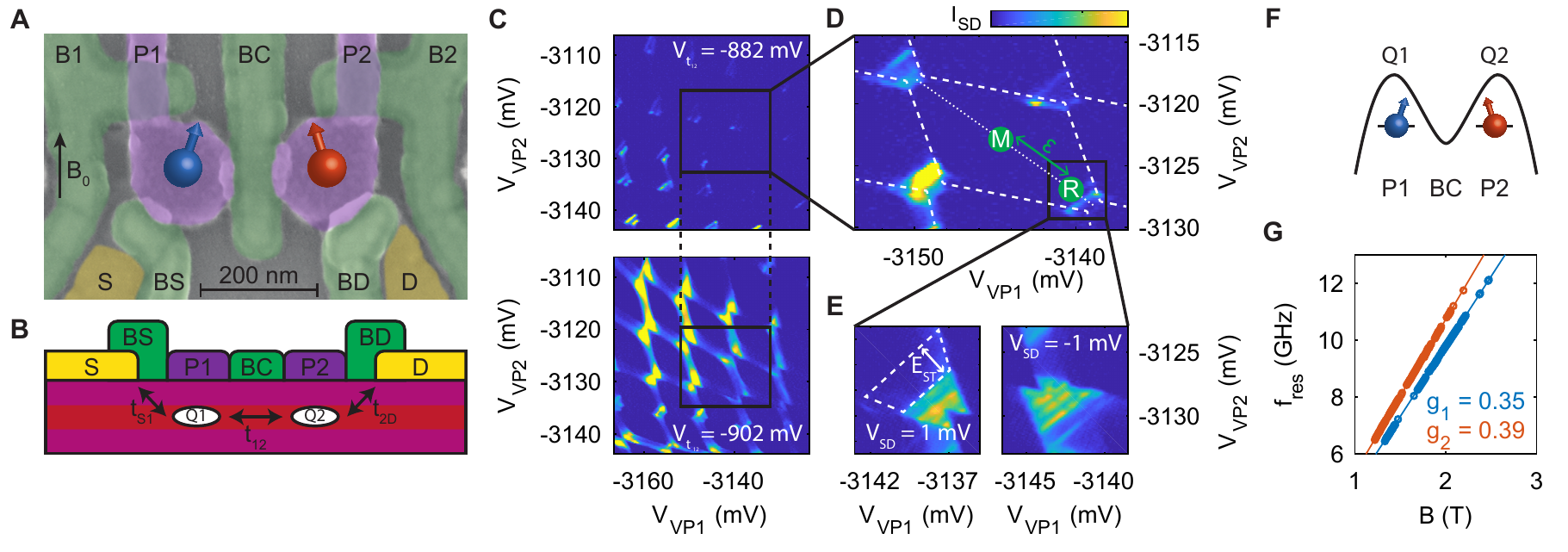}%
	\caption{
	\textbf{Fabrication and operation of a planar germanium double quantum dot.}
	(A) False-colored SEM image of the two qubit device. Two hole quantum dots are formed in a high-mobility Ge quantum well and controlled by the electric gates. The direction of the external field $B_0$ is indicated by the black arrow.
	(B) Schematic cross section of the system, where quantum dots are formed below plunger gates P1 and P2, while the different tunnelling rates can be controlled by barrier gates BS, BD and BC. 
	(C) Transport current through the double dot as a function of plunger gates voltage for low (top) and strong (bottom) interdot coupling, mediated by a virtual tunnel gate.
	(D) Charge stability diagram of the qubit operation point. The detuning axis $\varepsilon$ is indicated by the dotted line.
	(E) Transport current through the double dot as a function of plunger gate voltage for positive (left) and negative (right) bias. Pauli spin blockade becomes apparent from the suppression of the transport current for positive bias direction.
	(F) Illustration of the energy landscape in our double quantum dot system.
	(G) Resonance frequency of the two qubits as a function of the external magnetic field, showing the individual qubit resonances.}
	\label{fig:1}
\end{figure*}

Here, we make this step and demonstrate single and two qubit logic with holes in planar germanium. Fabrication is based on silicon substrates and standard manufacturing materials. We grow strained germanium quantum wells, measured to have high hole mobilities $\mu~>$~500.000 cm$^2$/Vs and a low effective hole mass $m_h$ = 0.09 $m_e$ \cite{hendrickx_gate-controlled_2018, sammak_shallow_2019}, and predicted to reach $m_h$ = 0.05 $m_e$ at zero density \cite{morrison_observation_2014,terrazos_qubits_2018}. This allows us to define quantum dots of comparatively large size and we find excellent control over the exchange interaction between the two dots. We operate in a multi-hole mode, significantly reducing challenges in tuning and characterization, and thereby advantageous for scaling. We exploit the spin-orbit interaction for qubit driving and perform single qubit rotations at frequencies exceeding 100 MHz. This advantage of fast driving becomes further apparent in coherently accessing the Hilbert space of a two-qubit system. For example, in silicon the execution of a CNOT gate implemented with an on-chip stripline has been shown using microsecond long pulses \cite{veldhorst_two-qubit_2015,huang_fidelity_2018} and this timescale is typically reduced to 0.2-0.5 microseconds by incorporating nanomagnets \cite{watson_programmable_2018,zajac_resonantly_2018}. Here, we demonstrate that the spin-orbit coupling of holes in germanium together with the sizable exchange interaction enables a CNOT within $t_\text{CX}$ = 75 ns.

A scanning electron microscope (SEM) image of the germanium two-qubit device is shown in Fig.~\ref{fig:1}A. In order to accumulate holes and define two quantum dots, the two circular plunger gates are set to negative potential ($V_\text{P1}, V_\text{P2} \approx -2$ V). The tunnel coupling between the dots $t_{12}$ and the tunnel couplings to the source and drain reservoirs ($t_{1\text{S}}$, $t_{2\text{D}}$) are controlled by the barrier gates BC, BS and BD, respectively. Working in a virtual gate voltage space (V$_\text{VP1}$, V$_\text{VP2}$, V$_{t_{1\text{S}}}$, V$_{t_{2\text{D}}}$, and V$_{t_{12}}$, respectively), we can independently tune these properties (see video mode operation in supplementary materials). We measure the transport current through the double dot system (Fig.~\ref{fig:1}C and \ref{fig:1}D) and for certain hole occupations we observe a suppression of the transport current for a positive bias voltage $V_\text{SD}=1$~mV, caused by Pauli-spin blockade (PSB) \cite{ono_current_2002}, see Fig.~\ref{fig:1}E. We make use of the blockade as an effective method for spin-to-charge conversion \cite{petta_coherent_2005}, as well as to initialize our two qubit system in the blocked $\ket{\downarrow\downarrow}$ ground state. 

\begin{figure*}%
	\includegraphics[width=18.5cm]{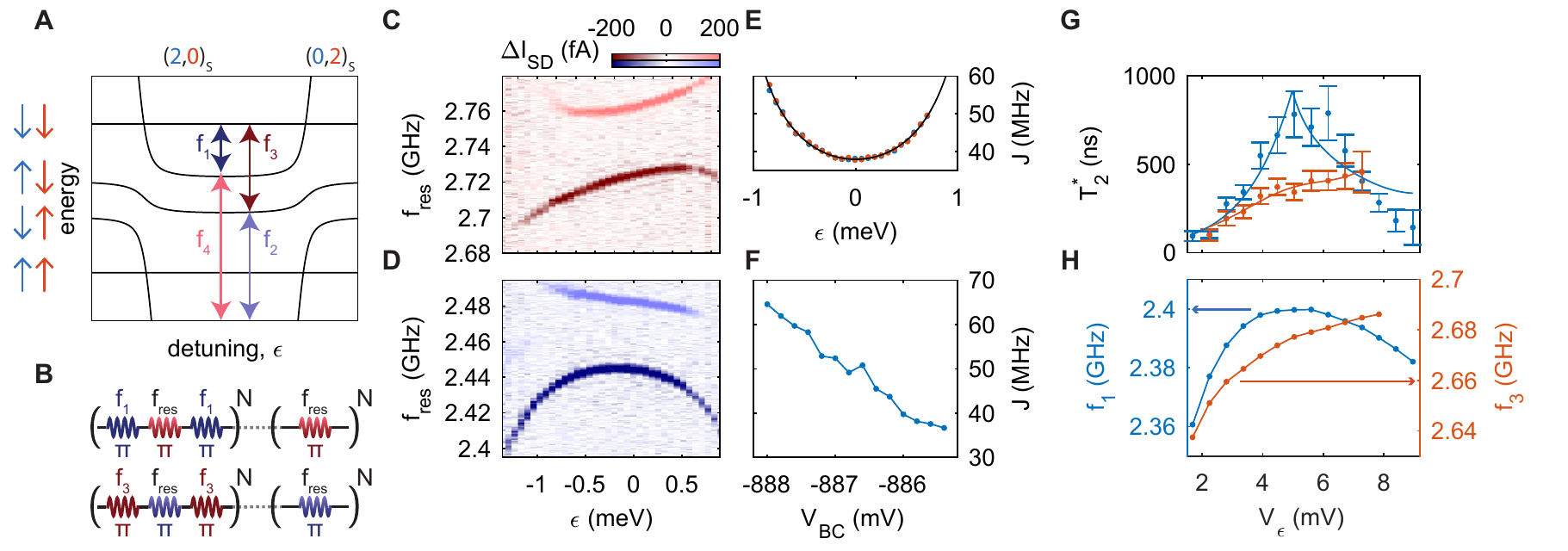}%
	\caption{
	\textbf{Tunable exchange coupling and operation at the charge symmetry point}
    (A) Illustration of the relevant energy levels in our hole double quantum dot with a finite exchange coupling $J$ between the dots. Four individual transitions can be driven, corresponding to the conditional rotations of the two qubit system.
    (B) Measurement pulse cycles used to map out the exchange splitting of Q1 (top) and Q2 (bottom). As a result of the demodulation of the alternating cycles, transition $f_{1,(3)}$ gives a negative signal and transition $f_{2,(4)}$ results in a positive signal.
    (C,D) EDSR spectra of Q1 (C) and Q2 (D) as a function of the detuning $\varepsilon$. The exchange splitting can be tuned to a minimum at $\varepsilon=0$ and increases closer to the $(m,n)-(m+1,n-1)$ and $(m,n)-(m-1,n+1)$ charge transitions.
    (E) Exchange interaction as a function of $\varepsilon$ as extracted from (C,D). Fitting the exchange coupling yields an interdot tunnel coupling $t_{12}=1.8$ GHz and charging energy $U=1.46$ meV.
    (F) The interdot tunnel coupling can also be controlled by gate BC. Changing the potential on this gate, while keeping $\varepsilon=0$, allows for a good control over the exchange interaction between the two qubits.
    (G) Coherence time $T_2^*$ of both qubits as a function of detuning voltage $V_\varepsilon$. When the slope of the resonance line is equal to zero, the qubit is expected to be in first order insensitive to charge noise. Solid lines indicate fits of the data to $\left(a\frac{\delta f_{res}}{\delta V_\varepsilon}+T_{0}\right)^{-1}$, with $\frac{\delta f_{res}}{\delta V_\varepsilon}$ the numerical derivative of the resonance line frequency as a function of detuning, $T_0$ the residual decoherence and $a$ a scaling factor. It can be observed that indeed $T_2^*$ is longest when the slope of the resonance line is closest to zero.
    (H) Resonance frequency of transition $f_1$ and $f_3$ as a function of detuning.
    }
	\label{fig:4}
\end{figure*}

Taking advantage of the strong spin-orbit coupling \cite{terrazos_qubits_2018, watzinger_germanium_2018}, we are able to implement a fast manipulation of the qubit states by electric dipole spin resonance (EDSR). We tune the device to a read-out point within the PSB-region (indicated by the label R in Fig.~\ref{fig:1}D) and apply an electric microwave excitation to gate P1. When the frequency of the microwave excitation matches the spin resonance frequency of either qubit, the PSB is lifted and an increase in the transport current can be observed. We extract the resonance frequency of each qubit as a function of external magnetic field strength $B_0$ and observe two distinct qubit resonance lines with $g$-factors $g_1 = 0.35$ and $g_2 = 0.38$ (Fig.~\ref{fig:1}G). The difference in g-factors between the two dots is likely caused by slightly different hole fillings and thus quantum dot orbitals. Due to the influence of spin-orbit coupling, a strong orbital dependence of the effective $g$-factor is typically expected in hole quantum dots \cite{liles_spin_2018,nenashev_wave_2003}. Furthermore the effective $g$-factor can be tuned electrically as a direct result of the SOC \cite{maier_tunable_2013} (see e.g. Fig. \ref{fig:4}C and \ref{fig:4}D), thereby guaranteeing independent control of the different qubits.

\begin{figure*}[ht]%
	\includegraphics[width=12cm]{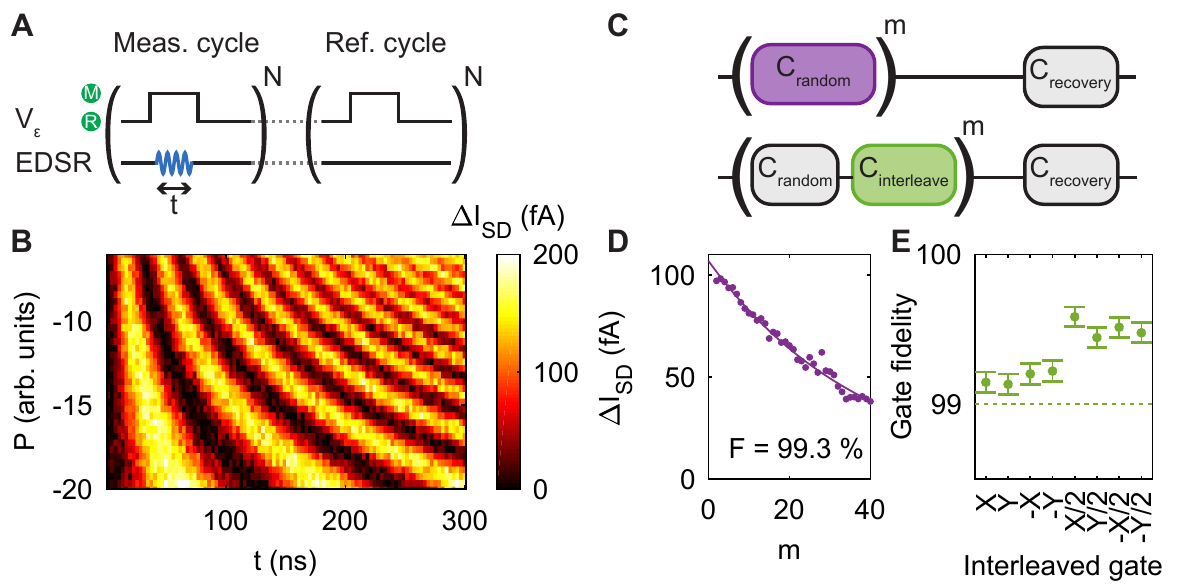}%
	\caption{
	\textbf{Coherent control and gate fidelity of a planar germanium qubit.}
	(A) Measurement sequence used for the Rabi driving measurements. Measurement cycli with EDSR pulses are alternated with reference cycli without a microwave tone, allowing for an efficient background current subtraction. Each cycle is repeated N times, such that measurement and reference cycles alternate at a typical lock-in frequency of $f_{meas}=89.75$ Hz (as discussed in the supplementary materials).
	(B) Colour map of the differential bias current $\Delta I_{SD}$ as a function of microwave pulse time $t$ and power $P$ and clear Rabi rotations on Q1 can be observed.
	(C) Schematic illustration of the (interleaved) randomized benchmarking sequence applied to Q1. 
	(D) Differential bias current as a function of the number of applied Clifford gates in the randomized benchmarking sequence on Q1. Q2 is used as an ancilla for the readout. The extracted control fidelity is $F_\text{C}=99.3\pm0.05$ \%.
	(E) Gate fidelities for the different $\pi$ and $\pi/2$ gates.}
	\label{fig:2}
\end{figure*}

To allow for coherent control of the isolated spin states, a two-level voltage pulse on gates P1 and P2 is used to detune the dot potentials and prevent tunneling to and from the dots during the manipulation phase (label M in Fig.~\ref{fig:1}D). We measure the averaged transport current over $N$ repeated pulse cycles and subtract a reference measurement to mitigate slow drifts in the transport current (cf. supplementary materials), as is indicated in Fig.~\ref{fig:2}A. The number of repetitions $N$ of each cycle is chosen to result in a lock-in frequency of $89.75$ Hz. 
During the readout (label R in Fig.~\ref{fig:1}E), no differential current is observed when the qubits are in their $\downarrow\downarrow$ ground state, while a signal of typically $\Delta I_{SD} \approx 0.3$ pA is measured for all other configurations and a total cycle length of $t_\text{cycle}=900$ ns. This is in good agreement with a bias current $\Delta I=2e/t_\text{cycle}=0.4$ pA, as expected for the random loading of a hole spin. After readout, the system is left in the blocking $\downarrow\downarrow$ state, serving as the initialization of our qubits.

We now operate the device in the single-qubit transport mode in an external field of $B_0=0.5$ T and use the second qubit as a readout ancilla. Coherent control over the qubit is demonstrated in a Rabi experiment, where the spin state of Q1 is measured as a function of microwave pulse length $t$ and power $P$, as shown in Fig.~\ref{fig:2}B. By increasing the power of the microwave pulse, we can reach Rabi frequencies of over a $100$ MHz, at an elevated $B_0 = 1.65$ T (as shown in the supplementary materials).

To evaluate the performance of our device, we implement randomized benchmarking of the single qubit Clifford group \cite{knill_randomized_2008} (Fig.~\ref{fig:2}C). The measured decay curve of the qubits state as a function of sequence length $m$ is shown in Fig.~\ref{fig:2}D, from which we can extract a single qubit control fidelity of $F_\text{C}=99.3\%$. Each data point is averaged over approximately 100000 repetitions of 1500 randomly drawn gate sequences. For each of these sequences, two recovery gates C$_\text{Recovery}$ are chosen such that the qubit is projected to the $\ket{\uparrow}$ and $\ket{\downarrow}$ state, respectively, realizing alternating cycles for lock-in detection. In Fig.~\ref{fig:2}E we show the gate fidelities for the different $\pi$ and $\pi/2$ gates as obtained by interleaved randomized benchmarking, where each randomly drawn gate is followed by the interleaved gate (cf. Fig.~\ref{fig:2}C). All individual gate fidelities are $F_\text{C}>99 \%$, with the infidelity for $\pi/2$ gates being approximately twice as low as for the $\pi$ gates, on account of the difference in pulse length.

We further characterize the individual qubits by measuring the coherence and relaxation times. We perform a Ramsey experiment, in which two $\pi/2$ pulses are separated by time $\tau$, during which the qubit will evolve as a result of the implemented detuning. We fit the decay of the observed oscillations to $\Delta I_{SD}~=~a\cos(2\pi f\tau+\phi)\exp(-(\tau/T_2^*)^{\alpha^*})$ and find a spin coherence time of $T_{2,Q1}^*~=~817$~ns and $T_{2,Q2}^*~=~348$~ns and decay coefficients of $\alpha^*_{Q1}=1.2\pm0.2$ and $\alpha^*_{Q2}=1.5 \pm 0.2$, for Q1 and Q2 respectively. These coherence times are a few times larger than reported for germanium hut wires \cite{watzinger_germanium_2018}. We note that due to the nature of the transport measurements, we cannot turn the coupling between the qubits fully off and thus a residual coupling of $J\approx 20$~MHz remains for these measurements, potentially limiting the single qubit performance. 

The spin coherence can be extended by performing a Hahn echoing sequence, consisting of $\pi/2$, $\pi$ and $\pi/2$ pulses separated by waiting times $\tau$. Fitting the observed decay as a function of the total waiting time $2\tau$ to a power law $\Delta I_{SD}~=a\exp(-(2\tau/T_2^*)^{\alpha^H})$, we find extended coherence times of $T_{2,Q1}^H~=~1.9~\mu$s and $T_{2,Q2}^H~=~0.8~\mu$s and decay coefficients of $\alpha^H_{Q1}=1.5\pm0.1$ and $\alpha^H_{Q2}=2.5\pm0.3$, for Q1 and Q2 respectively. Arsenic dopents in natural germanium attain values of $T_2=57~\mu$s and can reach long quantum coherence with $T_2=2T_1$ in isotopically purified samples. The difference in quantum coherence may have different origins, for example due to a difference in hyperfine interaction, in which case a dramatic improvement may be expected for isotopically purified germanium quantum dots \cite{itoh_high_1993,itoh_isotope_2014}.

Finally, we perform measurements of the spin relaxation times $T_1$ of the qubits by applying a $\pi$ pulse and waiting for time $\tau$ before performing readout. We observe spin relaxation times of $T_{1,Q1}=9~\mu$s and $T_{1,Q2}=3~\mu$s.  Furthermore, we find that these relaxation times increase exponentially when lowering the tunnel coupling between each qubit and its respective reservoir (data in supplementary materials) and longer relaxation times have been reported for germanium nanowires \cite{hu_hole_2012,vukusic_single-shot_2018}, giving good prospect for significantly increasing the relaxation time by closing the reservoir barrier during operation.

\begin{figure*}%
	\includegraphics[width=12cm]{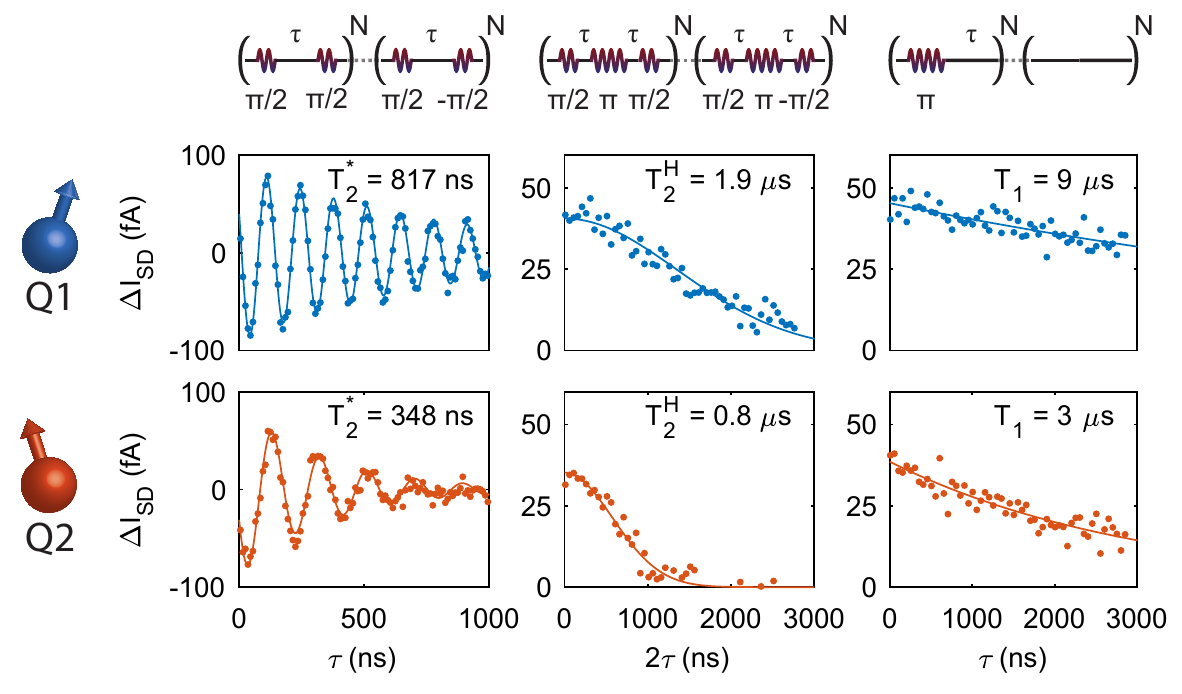}%
	\caption{
	\textbf{Relaxation, dephasing and coherence times.}
    Characteristic time scales of both qubits Q1 and Q2. For Q1 we find a coherence time $T_2^*=817$ ns (top left), which can be extended by applying a Hahn echo to $T_2^H=1.9~\mu$s (top middle). Furthermore, we observe a spin relaxation time $T_1=9~\mu$s (top right). For Q2 we observe a slightly shorter coherence time $T_2^*=348$ ns (bottom left). Applying a Hahn echo extends this coherence time to $T_2^H=0.8~\mu$s (bottom middle), and a spin relaxation time $T_1=3~\mu$s (bottom right) can be observed. 
    }
	\label{fig:3}
\end{figure*}

When the manipulation of both qubits is combined, the coupling of the two qubits (exchange interaction $J$) becomes apparent. As is illustrated in Fig.~\ref{fig:4}A, the resonance frequency of each of the qubits is shifted when the respective other qubit is prepared in its $\uparrow$ state. The strength of this interaction depends on the inter-dot tunnel coupling $t_{12}$ as well as the detuning $\epsilon$ of the dot potentials. By changing the depth of voltage pulse to point M (dashed line in Fig.~\ref{fig:1}E), we can map $J$ as a function of $\epsilon$. This is shown in Fig.~\ref{fig:4}C and \ref{fig:4}D, where the subtraction of two pulse sequences in the measurement (see Fig.~\ref{fig:4}B) results in a positive signal for the unprepared qubit resonances and a negative signal for the prepared states (cf. supplementary materials). As shown in Fig.~\ref{fig:4}E, the exchange coupling that is reflected in the frequency difference between the initial and prepared resonance positions, is very well described by a simple description using $J=4Ut_{12}^2/(U^2-(\alpha\epsilon-U_0)^2)$ \cite{russ_high-fidelity_2018,loss_quantum_1998}. Here, $U$ is the charging energy of the quantum dots, $\alpha = 0.23$ is the lever arm of P1 and P2, the interdot tunnel coupling is $t_{12}/h = 1.8$ GHz.
In addition, the strength of $t_{12}$ can be tuned using the central barrier BC (Fig.~\ref{fig:4}F). Here, we use a virtual gate voltage $V_{t_{12}}$, where $V_\text{BC}$ is set while compensating its influence on the dot potentials by appropriate corrections to $V_\text{P1}$ and $V_\text{P2}$ \cite{hensgens_quantum_2017, van_diepen_automated_2018}.
As a result of this full control over the coupling, we are able to operate the qubits at a mostly charge-insensitive point of symmetric detuning while choosing an exchange coupling strength large enough for rapid two qubit controlled rotations.

The advantage of this reduced sensitivity to detuning noise is demonstrated in Fig.~\ref{fig:4}G, where the dephasing time $T_2^*$ of both qubits is measured as a function of $\epsilon$. Here, $T_2^*$ directly reflects the slope of the frequency dependence of $f_{1,(3)}$ for Q1 and Q2, with the longest average phase coherence reached in the flat region $V_\epsilon\approx 6$ mV.

\begin{figure*}%
	\includegraphics[width=18.5cm]{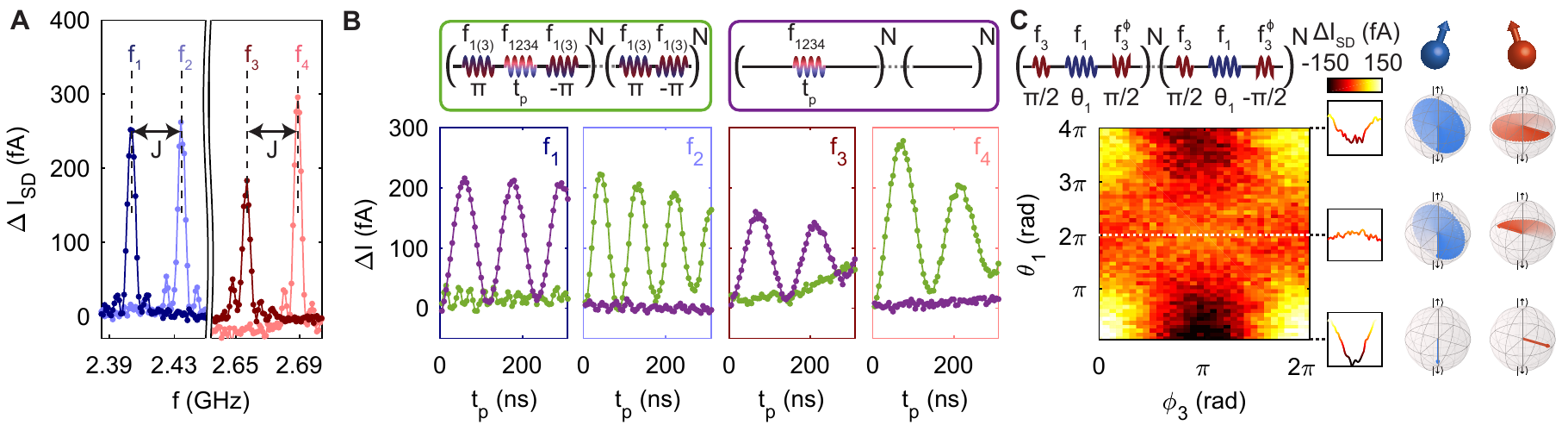}%
	\caption{
	\textbf{Fast two-qubit logic with germanium qubits.}
    (A) EDSR spectra of both qubits. Resonance peaks can be observed, corresponding to the four individual transitions indicated in Fig. \ref{fig:4}A. The peaks are power broadened and the line width is thus determined by the Rabi frequency.
    (B) Controlled qubit rotations can naturally be performed by selectively driving each of the four transitions. A CX gate is achieved when $t_{p}=t_\pi$ on $f_1$. A small off-resonant driving effect can be observed, which we mitigate by tuning $t_{\pi,\text{res}}=t_{4 \pi,\text{off-res}}$.
    (C) Colour plot of $\Delta I_{SD}$ as a function of Q1 CX-pulse length $\theta_1$ and the phase of the second $\pi/2$-rotation on Q2 $\phi_3$. Due to the $\theta/2$ Z-rotation on the control qubit, a $\pi$ phase shift can be observed on Q2 for a conditional $2\pi$ rotation on Q1 ($f_1$).
    }
	\label{fig:5}
\end{figure*}

The direct control over the tunnel coupling enables us to set the exchange interaction to a sizable strength of $J/h = 39$ MHz at the symmetry point, as demonstrated in Fig.~\ref{fig:5}A. We exploit this regime to obtain fast selective driving and operate in an exchange always-on mode. Full control is obtained by applying microwave pulses at the four resonant frequencies, while further gate pulses controlling $J$ are not needed. A pulse at a single resonance frequency will result in a conditional rotation of the target qubit, as we show in Fig.~\ref{fig:5}B. The slight off-resonant driving that can be observed on $f_{1\downarrow}$ is mitigated by choosing the driving speed such that $t_{\pi,\text{res}}=t_{4\pi,\text{off-res}}$. A fast CX-operation is thus achieved within $\tau_\text{CX,Q1}$ = 55 ns and $\tau_\text{CX,Q2}$ = 75 ns, with Q1 and Q2 as the target qubits respectively.

As a result of the pulsing, we observe a minor shift in the resonance frequency of both qubits. This was observed before in Si/SiGe quantum dots \cite{takeda_optimized_2018}, and we speculate this to be caused by a rectification of the AC microwave signal, leading to a slight change in the exchange interaction between the dots. We compensate the temporary change in resonance frequency by applying phase corrections to all following pulses (see supplementary materials). In Fig. \ref{fig:5}C we show the effect of a controlled rotation on the control qubit with applied phase corrections. We first apply a $X(\pi/2)$ pulse to Q2, followed by a $\theta_1$ conditional rotation of Q1. Finally, we apply a virtual $Z(\phi)$ gate by directly switching the phase on the final conditional $\pi/2$ pulse on Q2 ($f_3$). We observe larger phase rotations on Q1 after 0 and $4\pi$ rotations on Q2 as compared to a $2\pi$ rotation on Q2. This $4\pi$ periodicity is in agreement with fermionic statistics and suggests an echoing pulse correcting residual environmental coupling (see supplementary material). The full $\pi$ phase shift on Q2 for a conditional $2\pi$ rotation on Q1, as a result of the $\theta/2$ phase that is accumulated by the control qubit, demonstrates the application of a coherent CX gate.

The demonstration of a universal gate set with all electrical control and without the need of any microscopic structures provide great prospects to scale up spin qubits using holes in strained germanium. These quantum dots are furthermore contacted by superconductors \cite{hendrickx_gate-controlled_2018,hendrickx_ballistic_2019,vigneau_germanium_2019} that may be shaped into microwave resonators for spin-photon coupling, providing opportunities for a platform that can combine semiconducting, superconducting, and topological systems for hybrid technology with fast and coherent control over individual hole spins. Moreover, the demonstrated quantum coherence and level of control make planar germanium a natural candidate to engineer artificial Hamiltonians for quantum simulation going beyond classically tractable experiments.
\acknowledgements{We thank L.M.K. Vandersypen, S. Dobrovitski and J. Helsen for valuable discussions. We acknowledge support through a FOM Projectruimte of the Foundation for Fundamental Research on Matter (FOM), associated with the Netherlands Organisation for Scientific Research (NWO).}
\bibliography{spinqubit}
\end{document}